\documentclass[preprint2]{aastex}

\usepackage{natbib}
\usepackage{graphicx}
\usepackage{txfonts}

\slugcomment{Not to appear in Nonlearned J., 45.}

\shorttitle{ }
\shortauthors{Wakelam et al.}

\begin{document}

%% LaTeX will automatically break titles if they run longer than
%% one line. However, you may use \\ to force a line break if
%% you desire.

\title{A KInetic Database for Astrochemistry (KIDA)}

%% Use \author, \affil, and the \and command to format
%% author and affiliation information.
%% Note that \email has replaced the old \authoremail command
%% from AASTeX v4.0. You can use \email to mark an email address
%% anywhere in the paper, not just in the front matter.
%% As in the title, use \\ to force line breaks.

\author{V. Wakelam\altaffilmark{1,2}, E. Herbst\altaffilmark{3,4}, J.-C. Loison\altaffilmark{5,6}, I. W. M. Smith\altaffilmark{7}, V. Chandrasekaran\altaffilmark{5,6}, B. Pavone\altaffilmark{1,2},
N. G. Adams\altaffilmark{8}, M.-C. Bacchus-Montabonel\altaffilmark{9}, A. Bergeat\altaffilmark{5,6}, K. B\'eroff\altaffilmark{10}, V. M. Bierbaum\altaffilmark{11}, M. Chabot\altaffilmark{12},
A. Dalgarno\altaffilmark{13}, E. F. van Dishoeck\altaffilmark{14}, A. Faure\altaffilmark{15}, W. D. Geppert\altaffilmark{16}, D. Gerlich\altaffilmark{17}, D. Galli\altaffilmark{18}, E. H\'ebrard\altaffilmark{1,2}, F. Hersant\altaffilmark{1,2}, K. M. Hickson\altaffilmark{5,6}, P. Honvault\altaffilmark{19,20}, S. J. Klippenstein\altaffilmark{21}, S. Le Picard\altaffilmark{22}, G. Nyman\altaffilmark{23}, 
P. Pernot\altaffilmark{24}, S. Schlemmer\altaffilmark{25}, F. Selsis\altaffilmark{1,2}, I. R. Sims\altaffilmark{22}, D. Talbi\altaffilmark{26}, J. Tennyson\altaffilmark{27},
J. Troe\altaffilmark{28,29}, R. Wester\altaffilmark{30},  L. Wiesenfeld\altaffilmark{15}}

\altaffiltext{1}{Univ. Bordeaux, LAB, UMR 5804, F-33270, Floirac, France}
\altaffiltext{2}{CNRS, LAB, UMR 5804, F-33270, Floirac, France }
%\email{wakelam@obs.u-bordeaux1.fr}
\altaffiltext{3}{Departments of Physics, Astronomy, and Chemistry, The Ohio State University, Columbus, OH 43210 USA}
\altaffiltext{4}{Departments of Chemistry, Astronomy, and Physics,  University of Virginia, Charlottesville, VA 22904 USA}
\altaffiltext{5}{Univ. Bordeaux, ISM, CNRS UMR 5255, F-33400 Talence, France}
\altaffiltext{6}{CNRS, ISM, CNRS UMR 5255, F-33400 Talence, France}
\altaffiltext{7}{University Chemical Laboratories, Lensfield Road, Cambridge CB2 1EW, UK}
\altaffiltext{8}{Department of Chemistry, University of Georgia, Athens, GA 30602, USA}
\altaffiltext{9}{Universit\'{e} de Lyon (Lyon I), LASIM, CNRS-UMR5579, 43 Bd 11 Novembre 1918, 69622 Villeurbanne Cedex, France}
\altaffiltext{10}{Institut des Sciences Mol\'eculaires dÕOrsay, CNRS and Universit\'e Paris-Sud, 91405 Orsay cedex, France}
\altaffiltext{11}{Department of Chemistry and Biochemistry, Center for Astrophysics and Space Astronomy, University of Colorado, Boulder, CO 80309, United States}
\altaffiltext{12}{Intitut de Physique Nucl\'eaire d'Orsay,IN2P3-CNRS and Universit\'e Paris-Sud, 91406 Orsay cedex, France}
\altaffiltext{13}{Harvard-Smithsonian Center for Astrophysics, Cambridge, MA 02138, USA}
\altaffiltext{14}{Leiden Observatory, Leiden University, P.O. Box 9513, 2300 RA Leiden, The Netherlands}
\altaffiltext{15}{UJF-Grenoble~1/CNRS, Institut de Plan\'etologie et d'Astrophysique de Grenoble (IPAG) UMR 5274, Grenoble, F-38041, France}
\altaffiltext{16}{Department of Physics, University of Stockholm,Roslagstullbacken 21, S-10691 Stockholm}
\altaffiltext{17}{Technische Universit\"{a}t Chemnitz, Department of Physics, 09107 Chemnitz, Germany}
\altaffiltext{18}{INAF-Osservatorio Astrofisico di Arcetri, Largo E. Fermi 5, I-50125, Firenze, Italy}
\altaffiltext{19}{Laboratoire Interdisciplinaire Carnot de Bourgogne, UMR CNRS 5209, UniversitŽ de Bourgogne, 21078 Dijon Cedex, France} 
\altaffiltext{20}{UFR Sciences et Techniques, UniversitŽ de Franche-ComtŽ, 25030 Besanon Cedex, France}
\altaffiltext{21}{Chemical Sciences and Engineering Division, Argonne National Laboratory, Argonne, IL, 60439, USA}
\altaffiltext{22}{Institut de Physique de Rennes, Equipe Astrochimie Exp\'erimentale UMR 6251 du CNRS - Universit\'e de Rennes 1, B\^at. 11c, Campus de Beaulieu, 35042 Rennes Cedex, France}
\altaffiltext{23}{ Department of Chemistry, University of Gothenburg, 41296 Gothenburg, Sweden}
\altaffiltext{24}{ Laboratoire de Chimie Physique, UMR 8000, CNRS, Univ. Paris-Sud, F-91405 Orsay}
\altaffiltext{25}{ I. Physikalisches Institut, University of Cologne, ZŸlpicher Str. 77, 50937 Cologne, Germany}
\altaffiltext{26}{Universit\'e Montpellier II - GRAAL, CNRS - UMR 5024, place Eug\`ene Bataillon, 34095 Montpellier, France}
\altaffiltext{27}{Department of Physics and Astronomy, University College London, Gower St., London WC1E 6BT, UK}
\altaffiltext{28}{Institut f\"ur Physikalische Chemie, Universit\"at G\"{o}ttingen, Tammannstr. 6, D-37077 G\"{o}ttingen, Germany }
\altaffiltext{29}{ Max-Planck-Institut f\"ur Biophysikalische Chemie, Am Fassberg 11, D-37077 G\"{o}ttingen, Germany}
\altaffiltext{30}{Institut f{\"u}r Ionenphysik und Angewandte Physik, Universit{\"a}t Innsbruck, Technikerstra{\ss}e 25, A-6020 Innsbruck, Austria}

\begin{abstract}

We present a novel chemical database for gas-phase astrochemistry.  Named the KInetic Database for Astrochemistry (KIDA), this database consists  of gas-phase reactions with rate coefficients and uncertainties that will be vetted to the greatest extent possible.  Submissions of measured and calculated rate coefficients are welcome, and will be studied by experts before inclusion into the database.  Besides providing kinetic information for the interstellar medium, KIDA is planned to contain such data for planetary atmospheres and for circumstellar envelopes.  Each year, a subset of the reactions in the database (kida.uva) will be provided as a network for the simulation of the chemistry of dense interstellar clouds with temperatures between 10~K and 300~K.  We also provide a code, named Nahoon,  to study the time-dependent gas-phase chemistry of 0D and 1D interstellar sources.

\end{abstract}


\section{Introduction}

The interstellar medium (ISM) is composed of gas and dust under a wide range of physical conditions. The diffuse interstellar medium is characterized by very low gas densities ($\le 10^3$ cm$^{-3}$), temperatures between 30 and 100~K, high UV photon fluxes and chemical species mainly in atomic form (except for molecular hydrogen). Due to gravitational interactions and turbulence, this material slowly forms denser ($\sim 10^4$ cm$^{-3}$) and colder (10~K) regions through which  ionizing cosmic rays can penetrate but UV photons cannot. Over a period of about $10^5$-$10^6$ yr, these atoms and molecular hydrogen start a series of chemical reactions that synthesize molecules in the gas and on dust particles. Some of these molecules are large species, mainly in the form of unsaturated (H-poor) carbon-chains.  Due to perturbations, these cold and dense regions, called dark molecular clouds or cold cores, collapse to begin the different stages of star and planet formation; these stages comprise a prestellar core, a protostar, a star with a protoplanetary disk, and then possibly a planetary system.  As the physical evolution occurs, a chemical evolution also occurs in which the dominant larger molecules are no longer unsaturated carbon chains but more saturated organic and prebiotic species quite common on the earth.  At the end of their lives,  stars re-inject their material into the ISM, now enriched in heavy elements. This set of events is the life cycle of stars and interstellar matter.

The study of the composition and evolution of interstellar material is part of what we call astrochemistry. During the next ten years, many fascinating discoveries are expected in this field of research thanks in part to  a number of new observational instruments including the Herschel Space Observatory (HSO), the Atacama Large Millimeter/Submillimeter Array (ALMA),  as well as the expanded Very Large Array (eVLA), and the Stratospheric Observatory for Infrared Astronomy (SOFIA).  HSO and SOFIA are currently observing molecular spectra in the far-infrared,  a frequency range not accessible from the ground, whereas the ALMA interferometer, now only in its early science stage, will  give access to very high spectral resolution and sensitivity,  with broadband spectral coverage as well as spatial resolution never achieved before.  Many new molecules have been and will be detected, increasing our knowledge of the complexity of the ISM. Moreover, astrochemistry is important for the understanding of Sun-like star and planetary formation. Indeed,  protostellar collapse is governed by quantities such as the ionization fraction (directly related to the chemical composition), which couples the gas to the magnetic field. Finally, the evolution of the chemical composition of the gas, dust, and ice in the protoplanetary disk is a key process for the understanding of planetary formation, the fate of complex organics acquired in earlier stages, as well as the emergence of life.

In order to understand the formation and destruction of molecules observed in assorted sources in the ISM, astrochemists model the chemical evolution through reaction networks. In these models, the chemistry is described by processes of various types: ion-neutral and neutral-neutral reactions, dissociation and ionization by UV photons and cosmic ray particles, dissociative recombination, etc. 
The models compute the abundances as a function of time for up to four hundred gas phase species using more than four thousand reactions, starting from an initial composition.  Extensions that include isotopologues, anions, and large molecules contain many more reactions and molecules.  For dark clouds, the initial composition can consist of abundances measured in more diffuse regions, mainly atoms and molecular hydrogen.  Two main chemical databases of gas-phase chemistry currently exist in astrochemistry for the interstellar medium. The better known is the udfa database (previously called the UMIST database), created by Tom Millar (Queen's University of Belfast, Northern Ireland), first released in 1991 \citep{1991A&AS...87..585M,1997A&AS..121..139M,2000A&AS..146..157L,2007A&A...466.1197W}. Updated versions of the database have been released in 1995, 1999, and 2006 (http://www.udfa.net/). The second one is the OSU database, initially developed by Prasad and Huntress (1980), updated by  Chun Leung and Eric Herbst, and then revised over the last twenty years by Herbst and collaborators. Several versions dating back to 2003 are available at http://www.physics.ohio-state.edu/$\sim$eric/research.html. Although the two databases are similar, some significant differences exist because of occasionally different choices of which reactions to include;  udfa includes more studied reactions whereas OSU relies more heavily on unstudied radical-neutral reactions.  In addition, the two networks differ partially because of
\begin{itemize}
\item [1.] Different approaches to estimate the rate coefficients of reactions not studied in the laboratory,

\item [2.]	Different choices of experimental rate coefficients from uncritical compilations (the NIST database for instance \footnote{http://kinetics.nist.gov/kinetics/index.jsp}, and

\item [3.]	Different approximations regarding the temperature dependence of ion-polar neutral reactions.
\end{itemize}

Until its most recent version, which is useable up to 800 K, the OSU database had been more relevant for cold sources ($\sim$ 10~K) whereas udfa includes many reactions with energy barriers for hotter conditions ($>$ 100~K). Other accessible databases exist for particular cases such as Srates\footnote{http://kida.obs.u-bordeaux1.fr/uploads/models/Srates$_-$update.txt} for sulfur chemistry \citep{2004A&A...422..159W,2011A&A...529A.112W}, a database for shock models \citep{1993MNRAS.262..915P} and one for photon-dominated regions\footnote{http://pdr.obspm.fr/PDRDB/} \citep{2006ApJS..164..506L}. The existence of these diverse databases, which contain many unstudied reactions, and often lack references, is a barrier to the improvement of chemical models. We have therefore initiated a project of constructing a unique and critical database for gas-phase astrochemistry named the Kinetic Database for Astrochemistry\footnote{http://kida.obs.u-bordeaux1.fr} (KIDA). This database will include as many of the important gas-phase reactions for modeling various astronomical environments and planetary atmospheres as possible. Indeed, reaction networks relevant for planetary atmospheres require many of the same reactions as those in interstellar modeling, but over a different range of temperatures.

\section{General presentation of the database}

KIDA is designed in order to allow the efficient managing of data, user-friendly visualization of data and flexible downloading of lists of reactions. In addition to being able to consult the database, users can submit data to KIDA via an online interface. 

\subsection{Data on chemical species}

Users can find two different types of data in KIDA. Some data are related to chemical species and others are related to  chemical reactions. In the first type, one can find InChI codes, CAS registry numbers, electronic states, polarizabilities, dipole moments, and enthalpies of formation. InChI codes\footnote{http://www.iupac.org/inchi/} (for International Chemical Identifier) are character chains defined by IUPAC to identify any chemical species. The InChI code of OCS for instance is InChI=1S/COS/c2-1-3. These codes are needed for clear identification, especially in the context of interoperability of databases. Free software is available to draw the structure of the molecule from the InChI code. CAS registry numbers are numeric identifiers from the American Chemical Society. These numbers exist, however, only for chemical species used in industry.
% KIDA also contains some thermodynamical data such as enthalpy of formation, polarizabilities and dipole %moments.
Most of these data in KIDA have been collected from other databases such as the NIST Chemistry Webbook\footnote{http://webbook.nist.gov/chemistry/form-ser.html}, CCCBDB\footnote{http://cccbdb.nist.gov/}, and  \citet{2009ApJS..185..273W}. JPEG images of the structures of the species, drawn with the JMOL free software\footnote{http://jmol.sourceforge.net/index.fr.html}, are also displayed. 
 
\subsection{Rate coefficients}
 
For chemical reactions, rate coefficients are stored in the form of three parameters (A, B and C), which can be used to compute the rate coefficients using different formulae depending on the considered process and the range of temperatures desired. Rate coefficients define the intrinsic velocity of chemical reactions; the units in which they are expressed depends on the order of the reaction.  The overall rate of a reaction is given by the product of the rate coefficient and the densities of the reactants. There are nine classes of reactions in KIDA, which are listed and described  in Table~\ref{itype}.   The first three classes can be thought of as unimolecular processes induced by cosmic rays or photons; here the rate coefficient has units of s$^{-1}$, and reaction rates are computed by multiplying the rate coefficient by the density (in cm$^{-3}$) of the reactant. The next five classes involve bimolecular collisions; in these classes rate coefficients have units of cm$^3$~s$^{-1}$, and the rate of reaction is computed by multiplying the rate coefficient by the densities (in cm$^{-3}$) of the two reactants. The last type is composed of ter-molecular (three-body) association reactions, which are processes where two reactants associate to form a product that is stabilized by collision with a third species, which can be a reactant or an inert atom or molecule.   In the low-density limit, the rate of reaction is given by a rate coefficient in units of cm$^{6}$ s$^{-1}$ multiplied by the product of the densities of the three species involved in the reaction.  A more general rate law involves a more complex dependence on density, which eventually saturates into a bimolecular rate law \citep{1993JPCRD..22.1469A}. Three-body association reactions are effective in ``very dense'' sources by interstellar if not terrestrial standards ($> 10^{11}$ cm$^{-3}$) and are generally considered in the atmospheres of planets. This type of  reaction will be included later into KIDA. 

 The rate coefficients for bimolecular processes involving two heavy species or one heavy species and an electron  are each given a certain range of temperature in which the parameters are applicable.
By default, the rate coefficients are valid between 10 and 300~K.  For updated rate coefficients, the temperature range is given by the results of experiments or the estimates of experts.  Extrapolations outside this range of temperature are not recommended unless the rate coefficient is predicted to be totally independent of temperature, as occurs for the Langevin model of exothermic ion-non-polar-neutral reactions.  Temperature dependent rate coefficients, at a specific temperature $T$, can often be fit to  the Arrhenius-Kooij formula: 
\begin{equation}
k_{Kooij}(T) = A (T/300)^B \exp(-C/T) ~{\rm cm}^3~{\rm s}^{-1},
\end{equation}
where $A$, $B$, and $C$ are parameters.  Several sets of rate coefficients for the same reaction can exist for different ranges of temperature or even the same range. If this happens for identified key reactions, KIDA experts can give recommendations on which values to use (see section~\ref{recommendations}).

 Rate coefficients for unmeasured reactions between ions and neutral species with a dipole moment are computed using the Su-Chesnavich capture approach as discussed in \citet{2009ApJS..185..273W} and \citet{2010SSRv..156...13W}.  Here the rate coefficient $k_{\rm D}$ is expressed in terms of the temperature-independent Langevin rate coefficient $k_{\rm L}$  using two formulas, one for lower and one for higher temperatures.  The temperature ranges are divided by that temperature for which a parameter $x$, defined by the relation,
 \begin{equation}
 x = \frac {\mu_{\rm D}}{(2\alpha kT)^{1/2}},
 \end{equation}
 where (in cgs-esu units) $\mu_{\rm D}$ is the dipole moment and $\alpha$ the dipole polarizability,   is equal to two. The formulae for $k_{\rm D}/k_{\rm L}$ at the lower and higher temperature ranges are respectively
 \begin{equation}
 k_{\rm D}/k_{\rm L} = 0.4767x + 0.6200
 \end{equation}
 and
 \begin{equation}
 k_{\rm D}/k_{\rm L} = (x + 0.5090)^{2}/10.526 + 0.9754.
 \end{equation}
 At the lowest temperatures in the region of validity for equation (3) ($T \ge 10$~K), $k_{\rm D}$ goes as $T^{-1/2}$, whereas at the highest temperatures, $k_{\rm D}$ approaches the Langevin value.  The rate coefficients are valid in the so-called semiclassical region, before they level off with decreasing temperature near 10 K or below \citep{1996JChPh.105.6249T}.   

 Rate coefficients for photo-processes by UV photons, $k_{photo}$, are normally computed as a function of the visual extinction $A_v$: 
\begin{equation}
k_{photo} = A \exp(-C A_v)~{\rm s}^{-1}. 
\end{equation}
The rate  is obtained by multiplying $k_{photo}$ by the density of the reactant. The parameter $A$ for $k_{photo}$ represents the unattenuated photodissociation rate in a reference radiation field whereas $\exp(-C A_v)$ takes into account the continuum attenuation from the dust.  Since dust particles tend to attenuate light at shorter wavelengths more strongly, values of C are larger for species that have thresholds for photodissociation/photoionization farther into the UV \citep{1988rcac.book...49V}. Photoinduced rates  depend on the external UV radiation field assumed.  For example, the so-called average interstellar radiation field differs from the radiation field around a specific type of star.  It is possible in KIDA to specify which field will be used. The data included in KIDA have been computed for the interstellar radiation field  \citep{1978ApJS...36..595D}. Data computed with other filed will be included in the future.  

Cosmic rays, chiefly protons and alpha particles with energies above 1 MeV,  interact with gas-phase atoms and molecules in two ways. First, these high-energy particles can directly ionize a species. This direct process is dominant for H$_2$, H, O, N, He and CO.   Electrons produced in the direct process can cause secondary ionization before being thermalized.  The secondary process is taken into account in most treatments by a factor of 5/3 \citep{1999ApJS..125..237D,2012A&A...537A...7R} while the effect of ionization by alpha particles is taken account of by a separate factor of 1.8.  The total ionization rate coefficient  including these two factors  is computed by 
\begin{equation}
k_{CR,i} = A_i \zeta
\end{equation}
where $\zeta$ (s$^{-1}$) incorporates the flux of cosmic rays and secondary electrons and normally refers to atomic hydrogen \citep{1978ApJ...219..750C} . In the udfa, OSU, and KIDA networks, however, $\zeta$ refers to $molecular$ hydrogen ($\zeta_{H_2}$), and is approximately twice the value for atomic hydrogen.  The rate coefficients for different species can be obtained by including a factor $A_{i}$.

In addition to secondary ionization, the electrons produced in direct cosmic ray ionization can electronically excite both molecular and atomic hydrogen.  The radiative relaxation of H$_2$ (and H)  produces UV photons that ionize and dissociate molecules.  This indirect process  is known as the Prasad-Tarafdar mechanism \citep{1983ApJ...267..603P}. It represents a source of UV photons inside dense clouds. The rate coefficients of dissociation or ionization that proceed via excitation of molecular hydrogen are given by 
\begin{equation} 
k_{CRP,i} = \frac{A_i}{(1-\omega)} \frac{n(H_2)}{(n(H) + 2n(H_2))} \zeta_{H_2}
\end{equation}
 with $\zeta_{H_2}$ the total ionization rate for H$_{2}$,  $\omega$ the albedo of the grains in the far ultraviolet, $n(H)$ the density of atomic hydrogen and $n(H_2)$ the density of molecular hydrogen \citep[see also][]{1989ApJ...347..289G}. In the OSU network (and subsequently in KIDA), it is assumed that $\omega$ is 0.5 and that the abundance of atomic hydrogen is negligible compared to molecular hydrogen. The rate then simplifies to $k_{CRP,i} = A_i \zeta_{H_2}$. This of course is only valid in dense sources. Unlike the case for direct cosmic ray ionization, the factor $A_i$ can be quite large, often in the range 10$^{3} - 10^{4}$. The case of CO must be treated individually because of self-shielding \citep{1987ApJ...323L.137G}. 
 
Any data provided in KIDA is associated with a reference. This reference can be bibliographic, another database, or a datasheet, as discussed in Section~\ref{recommendations}. All the reactions from the low-temperature OSU database osu-01-2009 have been included, even reactions involving anionic species \citep{2008ApJ...685..272H}. Many updates concerning photoinduced rates are currently being undertaken based on results from \citet{2006FaDi..133..231V} and \citet{2008CP....343..292V}. Branching ratios for the dissociation of carbon clusters by both external UV radiation and cosmic ray-induced  photons   have been added based on \citet{2010A&A...524A..39C}.  Reactions necessary for primordial chemistry, collected by D. Galli, will be included before the end of 2012. Finally, many other additions have been made, in particular for neutral-neutral reactions, based on new experimental and/or theoretical results.  All the original bibliographic references listed in KIDA are given in an online appendix to this paper. 

\subsection{Recommendations concerning rate coefficients}\label{recommendations}

For all the rate coefficients in KIDA, a status is indicated. There are four possible recommendations, which are listed below:   

\begin{itemize}
\item [0.] Not recommended:   This status suggests strongly that the reaction not be included.
\item [1.] Unknown:  This status is mainly for reactions from other databases that have not been looked at by KIDA experts. 
\item [2.] Valid: This is a status meant for reactions that are associated with a bibliographic reference.
\item [3.] Recommended: Recommended rate coefficients are mainly deduced from a variety of sources (experimental or theoretical) but can also just be educated guesses. The recommended status must be justified by a datasheet and can evolve with time when new data are available. Datasheets are pdf files stored in KIDA\footnote{http://kida.obs.u-bordeaux1.fr/publications$\sharp$span$_-$datasheets} summarizing the information available on a reaction. These datasheets are inspired from the work done by the IUPAC subcommittee for Gas Kinetic Data Evaluation\footnote{http://www.iupac-kinetic.ch.cam.ac.uk/}. Datasheets, of which there are currently more than 40,  are signed by one or more authors and are reviewed internally by KIDA experts. 
\end{itemize}

\subsection{Uncertainties for rate coefficients}

The reliability of any reaction rate coefficient stored in KIDA is expressed in terms of an estimated uncertainty over the temperature range covered
by the recommendation. This uncertainty is specified by three fields, two of which are shown in Table~2: 
\begin{enumerate}
\item The type of the statistical distribution representing the uncertainty.
The preferred distribution, and the default, is the Lognormal (default
= logn). This choice enforces the positivity of the rate coefficient,
even for large uncertainty factors ($F_{0}\ge2$) \citep{Thompson1991,2009JPCA..11311227H}. Other distributions are implemented to
accommodate the most common uncertainty representations: Normal (norm),
Uniform (unif), and Loguniform (logu) (Table 2); 
\item The uncertainty parameter $F_{0}$. By default, it is the standard
geometric deviation for the Lognormal distribution, but has no default
value. It defines a standard ``$1\sigma$'' confidence interval around the reference
value $k_{0}$ \citep{Sander2011}, with the probability of assertion (Pr) defined by $\mathrm{Pr}(k_{0}/F_{0}\le k\le k_{0}*F_{0})\simeq0.68$.
For other distributions than the Lognormal, $F_{0}$ has different
meanings and units, as specified in Table 2;
\item The expansion parameter $g$ (default = 0, units: Kelvin). It is used
to parameterize a possible temperature-dependence of the uncertainty,
according to the formula $F(T)=F_{0}*\exp\left(g\left|1/T-1/T_{0}\right|\right)$
\citep{Sander2011}. In KIDA, the reference temperature $T_{0}$
is fixed at 300\,K \citep{2009JPCA..11311227H}. 
\end{enumerate}
More details on how to use, choose, and specify uncertainty representations
in KIDA, notably for data providers, is given in http://kida.obs.u-bordeaux1.fr/\\
help/helpUncert.html. 

\subsection{Member Area, Submitting Data to KIDA, and Downloading Lists of Reactions}\label{submit}

In order to be able to interact effectively with KIDA, users need to first register by providing an email address and a password. They will then have access to a member area where they can submit rate coefficients to KIDA experts and see the history of their downloads.   To improve the updating of KIDA, we have designed an online interface to upload data. Any user registered with KIDA can submit rate coefficients (with a temperature range of validity, an uncertainty and a reference). These data will then be sent to the "Super-Expert" of the type of reaction, who will then consult his/her experts to see if the data have been correctly included or not.   Lists of KIDA super-experts and experts with terms until 2013 are given in http://kida.obs.u-bordeaux1.fr/contact. 

It is also possible to download lists of reactions from KIDA using an online form\footnote{http://kida.obs.u-bordeaux1.fr/astrophysicist}. In this form, the user can select 1) the elements and/or species he is interested in, 2) the types of reactions and 3) the range of temperature. This last choice can be used, for instance, as a filter to remove some of the high temperature reactions if one is interested in modeling the cold interstellar medium. If no recommended rate coefficients exist in the full range of temperature asked by the user, several rate coefficients for the same reaction may be available in the downloaded file. These reactions are included in the readme file downloaded at the same time, and the user has to deal with them. If rate coefficients are defined over complementary ranges of temperature, the user may want to keep them and take into account this information in the model (see section~\ref{code}). Temperature ranges are indicated in the network along with the status of the KIDA recommendation (see section~\ref{recommendations}). In any case, we do not recommend the extrapolation of rate coefficients outside this range.  If this procedure must be undertaken, we recommend  instead using the results at the relevant extremum of temperature or consulting an expert or super expert from the KIDA lists.

For each reaction, KIDA stores up to three reactants, up to five products, the values of the kinetic parameters 
A, B and C, the uncertainty parameters $F_0$, $g$ and the type of uncertainty, the temperatures $T_{min}$ and $T_{max}$, which define the range of temperatures over which the rate coefficient is defined, the name of the formula used to compute the rate coefficient or the external UV field used to compute the photoinduced rates, the number of the reaction in the network, another number that represents a particular choice of A, B and C when more than one possibility has been included, and a numerical recommendation given by KIDA experts and discussed in the next section.
When there is no relevant temperature range,  -9999 and 9999 default values are used for the minimum and maximum temperatures.

\section{New Model for the Cold Dense Interstellar Medium}

\subsection{The Nahoon Public Code}\label{code}

The numerical code Nahoon is publicly available on KIDA to compute the gas-phase chemistry for astrophysical objects (http://kida.obs.u-bordeaux1.fr/models/).  The program is written in fortran 90 and uses the DLSODES (double precision) solver from the ODEPACK package (http://www.netlib.org/odepack/opkd-sum) to solve the coupled stiff differential equations. The solver computes the chemical evolution of gas-phase species at a fixed temperature and density (commonly referred to as zero-dimension, or 0D) but can be used in 1D if a grid of temperature, density, and visual extinction is provided. Nahoon is formatted to read chemical networks downloaded from KIDA. It contains a test to check the temperature range of the validity of the rate coefficients and avoid extrapolations outside this range. A test is also included to check for duplication of chemical reactions, defined over complementary ranges of temperature. In that case, Nahoon chooses the recommended rate coefficient. Some reactions, although important for the interstellar medium, are not included in KIDA such as H$_2$ formation on grains or recombination of cations with negatively charged grains. These reactions are, however, provided in a set of user instructions associated with Nahoon. This code represents a new version of the one that was put online in 2007 \citep{2008ApJ...680..371W}. Grains, both neutral and negatively charged, and electrons are considered as chemical species and their concentrations are computed at the same time as those of the other species.

\subsection{Gas-phase network for cold dense molecular clouds}

KIDA is a non-selective database. It may not be easy for all users to select the most appropriate list of reactions. For this reason, we will release every year a standard list of reactions for the dense interstellar medium between 10 and 300~K. This network will be available on the same page as Nahoon (http://kida.obs.u-bordeaux1.fr/models). 
To treat the formation of molecular hydrogen on granular surfaces, the OSU database uses the approximation from \citet{1971ApJ...163..155H}, which neglects any details of the process, and contains the assumption that the formation rate of
 H$_{2}$ is equal to 1/2 the collision rate of H atoms with grains \citep[see also][]{1984ApJS...56..231L}.  
  This approximation can induce some numerical problems during the computation because the elemental abundance of hydrogen is not rigorously conserved by the rather complex procedure used by the authors of the code.  Rather than improving the procedure, we use a simplified version of the more detailed approach in gas-grain models \citep{1992ApJS...82..167H}.   In particular, we have divided the H$_2$ formation into two reactions and included another species JH, which refers to hydrogen atoms on grains.  The first reaction is the sticking of atomic hydrogen onto grains (neutral or negatively charged): H $\rightarrow$ JH.

 The rate of this reaction (in cgs units of cm$^{-3}$~s$^{-1}$) is
\begin{equation}
R_{H, acc} = \pi a^2 \sqrt{\frac{8 k T}{\pi m_H}} n_d n(H)
\end{equation}
with $a$ the radius of grains (in cm), $k$ the Boltzmann constant, $m_H$ the atomic mass of hydrogen, $n_d$ the total density of grains, $T(K)$ the temperature, and $n(H)$ the density of atomic hydrogen.  
 For 0.1~$\micron$ radius grains, the rate is $7.92\times 10^{-5} \sqrt{T/300} n_d n(H)$.

The second reaction is the formation of H$_2$ on grains: JH + JH $\rightarrow$ H$_2$. We assume here that H$_2$ formed on grains is directly released in the gas phase. Since desorption of JH is not included, the process has an efficiency of 100\%. The rate of this reaction is  given by the expression
\begin{equation}
R_{H_2, form} = \frac{ \nu_0 \exp{(-E_b/kT)}}{N_s n_d}n(JH)^2
\end{equation}
with $N_s$ the number of binding sites on a grain, $\nu_0$ the characteristic or trial vibrational frequency, and
 $E_b$ the potential energy barrier between two adjacent surface potential wells. 
 
  A useful approximate formula is 
 $\nu_0 = \sqrt{2 n_s E_D k_B/ (\pi m_H)}$ \citep{1987ppic.proc..333T},  where $n_s$ is  the surface density of sites on a grain in cm$^{-2}$, and $E_D$ the adsorption energy of H in erg. The reader can find a detailed explanation in \citet{1992ApJS...82..167H} and \citet{2010A&A...522A..42S}.  Assuming grains of 0.1~$\micron$ radius, the number of sites per grain of $10^6$ \citep{1992ApJS...82..167H}, $E_D/k$ = 450~K,  and $E_b/k$ = 225~K,
$n_s$ is $7.96\times 10^{14}$ cm$^{-2}$, $\nu_0$ is $4.32\times 10^{12}$~s$^{-1}$ and the rate of H$_2$ formation is $8.64\times 10^{6} \exp{(-225/T)} n(JH)^2/n_d$ cm$^3$ s$^{-1}$.  See \citet{2006A&A...457..927G} for a discussion concerning the values of  $E_D$ and $E_b$, which depend upon the surface. The value of $n_{d}$ should obey the constraint that $\pi a^2$ $n_{d} \approx 1-2 \times 10^{-21}n_{\rm H}$, so that here $n_{d} \approx 5 \times 10^{-12} n_{\rm H}$, where $a$ is the granular radius and $n_{\rm H}$ is the total hydrogen nuclear density.

The second class of processes involves electron attachment to grains and the recombination of cations with negatively charged grains. Rate coefficients for such processes can be obtained in the ballistic approximation from the following equation:
\begin{equation}
k_{coll} = \pi a^2 \sqrt{\frac{8 k T}{\pi m_i}}
\end{equation}
where 
$m_i$ is the mass (g) of the impacting species (electrons or ions). We make the approximation that the mass of the impactor is negligible compared to the grain mass, so that the reduced mass is then approximately the mass of the lighter collider. For collisions between two charged species (cations and negatively charged grains), this rate can be multiplied by $(1 + \frac{Z e^2}{a k T})$ to account for the Coulomb attraction, with $e$ the elemental electronic charge and Z the number of negative charges on the grain \citep{1978MNRAS.184..227W}.  It is typically assumed, however, that,  in dense clouds, grains can be neutral or singly negatively charged only, reflecting the possibility that the trajectory of an electron nearing a negatively-charged grain is repulsive.  Following \citet{1984ApJS...56..231L}, we neglect any surface electronic charge, based on the assumption that the negative charge/charges on a grain remain localized so the additional factor does not play an important role in the dynamics of the recombination with cations. If the negative charges on a grain were delocalized, then the Coulomb interaction should be included. We have listed in Table~\ref{grain_reac} the minimum set of reactions with grains that should be included in a network for dense clouds. The rate coefficients (cm$^3$~s$^{-1}$) for these processes are computed from the A and B listed in the table using the equation $k = A (T/300)^B$, where A  is obtained for a grain radius of 0.1~$\micron$ and $B = 0.5$.  For grains of different radii, the ballistic rate coefficients simply scale as the square of the radius.  The referee has noted correctly that reactions involving gaseous ions and neutral grains are not included in the KIDA network.   They are not included because the only reactive gas-grain processes  in the network involve charge neutralization of negatively-charged grains.

In the interstellar medium, molecules can be dissociated by absorption of UV photons via electronic continua or through lines. The photodissociation of the dominant species H$_2$ and CO is dominated by line absorption (although the mechanisms are different) so that H$_2$ and CO self-shielding is an important process. To avoid a detailed treatment of radiative transfer, we adopted the approximation from \citet{1996A&A...311..690L}, which gives H$_2$ and CO photodissociation rates as a function of $A_v$, as well as the H$_2$ and CO column densities. To include self-shielding, the user of Nahoon has to set the conditions at the border of the cloud, including the abundances of H$_2$ and CO even when working in 0D. In 1D, a plane parallel geometry is assumed and the  UV photons enter the cloud from one direction.   Photodissociation rates at a spatial point are then computed using the column densities of H$_2$ and CO  according to the computed abundances and physical structure in front of this point. The visual extinction is then computed using the H$_2$ column density and the standard ratio between this column density and visual extinction. This geometry is not correct for clouds where UV photons can penetrate from both sides of the cloud.  Note that new shielding functions have been published by \citet{2009A&A...503..323V} and will be included in Nahoon in the future.

In addition to these processes, which are downloaded  with the new Nahoon, we have downloaded our first annual network of gas-phase chemical reactions necessary to compute the chemistry in the interstellar medium where the temperature is between 10 and 300~K and where surface chemistry is not critical. Approximations for the depth-dependent photodissociation rates are valid for $A_v$ above 5 \citep{1988rcac.book...49V}. The full network, designated kida.uva.2011, and the associated list of species are available at 
 http://kida.obs.u-bordeaux1.fr/models. This list contains 6090 reactions involving 474 species, including neutral and negatively charged grains, as well as electrons.  Elements that are included are H, He, C,  N,  O,  Si, S,  Fe, Na, Mg, Cl, P and F. 
 Anionic chemistry from \citet{2008ApJ...685..272H} has also been included. For temperatures in the range 300 K - 800 K, modelers can use the high-temperature OSU network osu-09-2010-ht, found and discussed on the URL http://www.physics.ohio-state.edu/$\sim$eric/research.html.

\subsection{Using uncertainties for rate coefficients in models}

Using uncertainty information in chemical modeling is vital, notably for the extreme environments targeted by KIDA, where many reaction
rate coefficients are known with little accuracy, either estimated or extrapolated. Uncertainty management has two goals:
\begin{enumerate}
\item Uncertainty Propagation: to estimate the precision of the model
outputs; and 
\item Sensitivity Analysis: to identify key reactions, i.e. those contributing notably to the uncertainty of model outputs and for which better experimental
or theoretical estimations are needed \citep{2010P&SS...58.1555D}. 
\end{enumerate}
Improvement of data in KIDA is based on an iterative strategy involving uncertainty propagation and sensitivity analysis \citep{2010SSRv..156...13W}.

\paragraph{Monte Carlo Uncertainty Propagation (MCUP).}

The simplest way to implement these methods is through Monte Carlo sampling \citep{Thompson1991,2008P&SS...56..519D,2005A&A...444..883W,2007P&SS...55..141C,2010SSRv..156...13W}. Random draws for a rate coefficient at  any temperature are generated using the formulae in Table~\ref{dist}. For MCUP, the code is run for $N$ random draws of the $M$ rate constants of the chemical scheme \{$k_{i}^{(j)};\, i=1,M;\, j=1,N$\}
(all parameters are perturbed simultaneously), and the $N$ sets of outputs are stored for further statistical analysis (mean value, uncertainty
factor, input/output correlation...).

\paragraph{Sensitivity Analysis.}

A convenient way to perform sensitivity analysis in the MCUP setup is to calculate the correlation coefficients between the inputs ($k_{i}$) and outputs
of the model. Large absolute values of correlation coefficients reveal strong influences of inputs on outputs \citep{2010SSRv..156...13W, 2010P&SS...58.1555D}. Variations of this method include using rank correlation coefficients, or the logarithm of inputs and/or outputs \citep{Helton06, Saltelli04}.

Monte Carlo simulations have been implemented in the Nahoon code bundle, and an IDL procedure has been provided to compute the standard deviation for the predicted abundances of species due to uncertainties in reaction rate coefficients. 

\subsection{Limitations}

 The interstellar chemical model defined by both the numerical program Nahoon and the osu.uva.2011 network takes into account a limited number of processes, which makes its use valid only over a limited range of temperature, density, and visual extinction. This model is meant to compute the gas-phase chemistry in regions where the visual extinction is larger than $\approx 3-4$. For smaller $A_v$, detailed PDR codes, such as the Meudon code \citep{2009A&A...505.1153L}, are more appropriate. Also, this model does not take into account positively charged grains, which play a major role at low extinction. The temperature limit is set by the network to the range 10 to 300~K, as already discussed.  For higher temperatures, one should adapt an expanded gas-phase network developed by \citet{HaHeWa2010}, which can be utilized up to gas temperatures of 800 K.  
The accretion time scale onto grains of radius $0.1\mu$ is roughly $10^{9}/n_{\rm H}(cm^{-3})$ yr.  At times  larger than this, the gas-phase can be significantly depleted of heavy species if the temperature lies below the 
 the evaporation temperature of easily vaporized species such as CO and CH$_{4}$.  At lower temperatures, non-thermal desorption mechanisms, such as photodesorption, can still maintain a residual gas phase.   Models including interactions with grains and grain surface processes \citep{2007A&A...467.1103G} are, however, based on many uncertain processes. In addition their use and the analysis of the results are more difficult. For this reason, pure gas-phase models can still be useful even for dense cloud conditions but users have to be careful when they compare their results with observations. 

The computation of H$_2$ formation on grains and the recombination of negative grains with cations, presented here, depend on the size of the grains. We have computed the rate coefficients for these processes for a single grain size of 0.1~$\mu$m, which with a typical grain density of 2.5 gm/cm$^{3}$, can explain observed extinctions with a gas-to-dust mass ratio of about 0.01, a reasonable value considering heavy elemental depletions in diffuse clouds \citep{2005pcim.book.....T}.  The effects of using a detailed grain size distribution on the rates of surface processes and on the chemical results for a full gas-grain model can be found in \citet{2011ApJ...732...73A}; these authors included grains large enough to have a constant rather than stochastic temperature and used a variety of size distributions \citep{2001ApJ...548..296W,1977ApJ...217..425M}.  They found little effect on the overall chemical results for a cold core with the OSU gas-grain network.   The relationship between granular size and rates of assorted processes is a complex one, because the rates can depend upon a number of  parameters in addition to surface area, such as whether the system lies in the accretion or diffusion limit and, for cold dense cores, the depth of the ice mantle.

The temperature dependence of the rate of H$_{2}$ formation on grains is a rather complex issue, and depends strongly on how rough the dust surface is and the availability of chemisorption sites, both of which can allow the process to continue up to much higher temperatures than the standardly-assumed Langmuir-Hinshelwood mechanism based on physisorption on smooth surfaces \citep{2005MNRAS.361..565C,2010ApJ...715..698C}.  The net result is that for systems with both physisorption and chemisorption sites, the efficiency of H$_{2}$ formation can remain high at all relevant interstellar dust temperatures, although this result is not guaranteed.  For small particles, in which the temperature is stochastic, the situation is more complex but has also been studied via Monte Carlo techniques \citep{2006MNRAS.367.1757C}.  For dust particles with well-defined temperatures up to 50 K, the physisorption mechanism assumed here should be adequate for rough particles.

%is the radius corresponding to the effective cross sections calculated from the grain size distribution from \citet{1977ApJ...217..425M} (IS IT UNDERSTANDABLE? IS IT CORRECT? DO YOU HAVE A REFERENCE FOR THAT? I HAVE THE IMPRESSION THAT THE REFEREE DOES NOT AGREE WITH THIS NUMBER.). Considering several sizes or even a distribution of sizes for the computation of the rate coefficients of Eqs. 8 and 10 may be more correct. SHOULD WE INCLUDE HIS COMMENT ON GAS-GRAIN MODELS? IT DOES NOT REALLY CONCERNS US. CAN WE SAY THAT THESE PROCESSES ARE NOT VERY IMPORTANT EVEN IN OUR MODEL?

Finally, despite their possible impact on  the interstellar chemistry of cold dense cores, as shown by \citet{2008ApJ...680..371W}, Polycyclic Aromatic Hydrocarbons (PAHs) are not included in this model because of the strong uncertainties that still remain (1) on the size and abundance of these species in dense regions and (2) on their reactions with other species. 

%The fact that their chemistry depends on their size (and structure), and the probable variety of sizes and structures in the ISM, makes their inclusion in chemical models difficult. The role that they play on the ionization fraction cannot, however be ignored, especially in regions where these PAHs have been observed.}

\subsection{Comparison with other networks}\label{comp}

\begin{figure*}
%\epsscale{.80}
\plotone{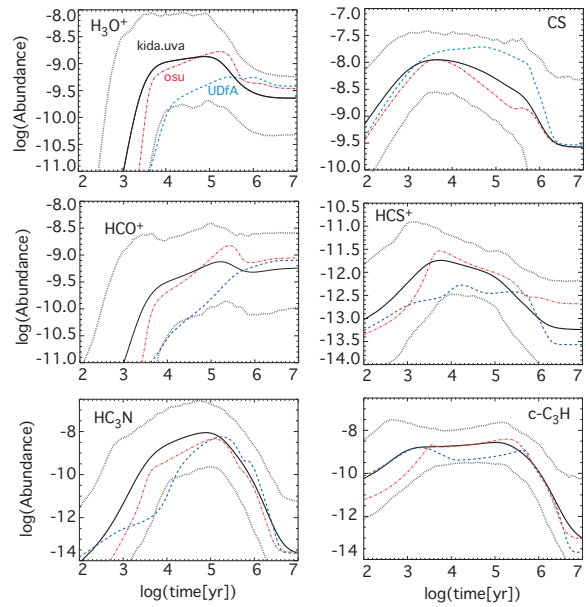}
\caption{Abundances as a function of time for selected species computed for dense cloud conditions (see section~\ref{comp}) using three different chemical networks: udfa06  (blue dashed lines),  osu-01-2009  (red dashed-dotted lines) and kida.uva.2011 (black lines). Dotted curves represent 2$\sigma$ error bars on the abundances computed with the kida.uva.2011 network using uncertainty propagation in the reaction rate coefficients. \label{fig1}}
\end{figure*}

To compare the abundances computed with the kida.uva.2011 network and other public networks, we have considered the case of dark molecular clouds using Nahoon in 0D. The temperature is set to 10~K, the proton density to $2\times 10^4$~cm$^{-3}$, the total H$_2$ cosmic ray ionization to $1.3\times 10^{-17}$~s$^{-1}$ and the visual extinction to 30. Initial abundances are taken from \citet{2010A&A...517A..21W}.
The results for selected molecular abundances, as computed with the kida.uva.2011 list of reactions,  are displayed in Fig.~\ref{fig1} with the results of the uncertainty propagation of rate coefficients.
For comparison, abundances computed using the udfa06 network  (http://www.udfa.net/) and osu-01-2009 network (http://www.physics.ohio-state.edu/$\sim$eric/research.html) networks are also included. Contrary to the publicly available udfa06 network,  the OSU network used also contains anions.

The particular molecular abundances plotted have been selected because they show some variation in their abundances with the database used. The OSU and  kida.uva networks lead to similar results since KIDA is mainly based on the OSU compilation although a number of updates have been performed.  Abundances computed with udfa06 on the contrary can be different,  as shown especially for CS and the cations in Fig.~\ref{fig1}.  Model predictions done with the different networks are however within the 2$\sigma$ error bars computed from the uncertainty propagation of rate coefficients, which are typically an order-of-magnitude in both directions.

\section{Conclusions}

We have presented in this article a new chemical database for gas-phase astrochemistry named KIDA (for KInetic Database for Astrochemistry).  KIDA can become a useful tool for physical chemists to circulate their relevant work for astrochemistry,  especially their most up-to-date data. In addition to kinetic data for the chemistry in the interstellar medium, KIDA will soon contain kinetic data for planetary atmospheres (other than the Earth). Kinetic data for carbon-rich and oxygen-rich circumstellar shells are also planned for the future.  State-to-state reactions, excited species, isotopologues and ortho and para forms will be considered for addition in KIDA.  The quality of the data present in KIDA will be improved over the years by KIDA experts and will be preserved thanks to the detailed information stored in KIDA. The visibility of KIDA will be improved by its interoperability with the Virtual Atomic and Molecular DataCenter \citep[][http://www.vamdc.eu/]{2010JQSRT.111.2151D} and the Europlanet Research Infrastructure (http://www.europlanet-eu.org/outreach/).  In addition to the rate coefficients, we are providing the community with a public fortran code, able to read networks downloaded from KIDA, called Nahoon, which can calculate the evolution of the gas-phase composition for 0D and 1D sources in the ISM and an extensive list of chemical reactions for the dense ISM with temperatures between 10 and 300~K.  Neither KIDA nor Nahoon treats the interaction of the gas-phase with interstellar grains, except for the formation of molecular hydrogen and recombination of the major cations with negatively charged grains. Omission of these processes is not realistic in the context of interstellar chemistry but the numerical and chemical complexity, and additional uncertainties brought by the treatment of surface chemistry, make pure gas-phase models still needed. In addition, gas-grain models  also rely on the quality of gas-phase rate coefficients.

\acknowledgments

The KIDA team acknowledges various sources of funding, which have allowed us to develop this database.  These sources include The University of Bordeaux, The ``Institut de Physique Fondamentale de Bordeaux", the CNRS/INSU (PCMI, PNP and ASOV), The Observatoire Aquitain des Sciences de l'Univers. We acknowledge support from the European Research Council (ERC Grant 209622: E$_3$ARTHs), the Agence Nationale de la Recherche (ANR-JC08$_-$311018: EMA:INC), the European program Astronet (CATS project), and the European Seventh Framework Programme (FP7: VAMDC and EUROPLANET projects). VAMDC is funded under the ÒCombination of Collaborative Projects and Coordination and Support ActionsÓ Funding Scheme of The Seventh Framework Program (Call topic: INFRA-2008-1.2.2 , Grant Agreement number: 239108).  The EUROPLANET RI (Research Infrastructure) is funded under the Grant Agreement number: 228319. The International Space Science Institute provided some of the KIDA team members with the opportunity to start this project in the context of the international team with the title "New generation of databases for interstellar chemical modeling in preparation for HSO and ALMA". VW and IWMS thank the Royal Society for its financial support of their collaboration. VW is grateful to all the participants of the project, who helped design the data model of the database. VW also thanks Franck Le Petit for helpful discussions on the rate coefficients for photodissociation induced by cosmic rays.  EH thanks the National Science Foundation (US) for his program in astrochemistry and NASA for the study of pre-planetary matter. NGA thanks NASA for support of his experimental program. SJK acknowledges support through NASA -PATM grant number NNH09AK24I. The anonymous referee is thanked for his/her useful comments, which helped to improve the quality of the paper.

\appendix

\section{Appendix}

\subsection{KIDA references}

Original references for kinetic data in KIDA are: \citet{Adams1976,Adams1984,Amano1990,Anicich1986, Anicich1976,Antipov2009,Atkinson2004,Bacchus-Montabonnel2008,Barckholtz2001,
Barlow1987,Baulch2005,Becker1989,Becker2000,Bedjanian1999,Bergeat2009,Berkowitz1986,Berteloite2010a,Berteloite2010b,Bettens1999,Bettens1995,Bohme1988,Brown1983,Brownsword1997,Brownsword1996,Canosa1997,Carty2006,2010A&A...524A..39C,CostesM.2009,Daugey2005,Davidson1990,Deeyamulla2006,Defrees1989,Derkatch1999,Dorthe1991,Duff1996,Ehlerding2004,Eichelberger2007,Fehsenfeld1975,Gamallo2006,Geppert2004a,Geppert2004b,Geppert2000,Grebe1982,1989ApJ...347..289G,Gu2006,HaolinSun2004,2008ApJ...685..272H,Herbst1989a,Herbst1985,Herbst1987,Herbst1989b,Herbst2000,Jensen2000,Kaiser2002,Kalhori2002,Lee1978,Lilenfeld1977,LiqueF.2009,Loison2009,Marquette1988,Martinet2004,Martinez2008,Martinez2010,Mayer1967,McGowan1979,Mebel2002,Messing1980,Millar1990,Montaigne2005,Neufeld2005,Ozeki2004,Parisel1996,Prasad1980,Rui-PingHuo2011,Sablier2002,Sander2011,Shin1986,Sims1994,Smith1981,Smith1989,Smith2001,Snow2009,Stevens2010,Talbi2009,Thorne1983,Thorne1984,Tuna2008,2006FaDi..133..231V,1988rcac.book...49V,VanHarrevelt2002,2008CP....343..292V,Vazquez2009,Viggiano2005,Vinckier1979,Vroom1969,Wakelam2009,Wennberg1994,2009ApJS..185..273W,Zabarnick1989}.

\bibliographystyle{apj}

\bibliography{aamnem99,kida,kida_all}

\clearpage

\begin{deluxetable}{lll}
\tabletypesize{\scriptsize}
\rotate
\tablecaption{Types of reactions defined in KIDA\label{itype}}
\tablewidth{0pt}
\tablehead{
\colhead{Type of reaction} & \colhead{Abbreviation} & \colhead{Description}}
\startdata
Direct cosmic ray processes &	CR 	&Dissociation or ionization of species due to direct collision with cosmic ray particles.\\
Photo-processes induced by cosmic rays (secondary photons) &	CRP 	&Dissociation or ionization of species due to UV photons emitted following H$_2$ excitation.\\
Photo-processes 	&Phot &	Dissociation or ionization of neutral species by UV photons with a standard interstellar UV field.\\
Bimolecular reactions &	Bimo &	Neutral-neutral (A + B $\rightarrow$ C + D), ion-neutral (A$^+$ + B $\rightarrow$ C$^+$ + D, A$^-$ + B $\rightarrow$ C$^-$ + D)\tablenotemark{a},\\ 
& & anion-cation (A$^+$ + B$^-$ $\rightarrow$ C + D) reactions and dissociative neutral attachment (A + B $\rightarrow$ AB$^+$ + e$^-$)\\
Charge exchange reactions &	CE & 	A$^+$ + B $\rightarrow$ A + B$^+$ and A$^+$ + B$^-$ $\rightarrow$ A + B \\
Radiative associations &	RA &	Association reactions between two species (neutral or ionized) stabilized by the emission \\
& & of a photon (A + B $\rightarrow$ AB + photon or A$^+$ + B $\rightarrow$ AB$^+$ + photon).\\
Associative detachment &	AD &	Association of a neutral species and an anion, resulting in the ejection of the extra electron (A$^-$ + B $\rightarrow$ AB + e$^-$).\\
Electronic recombination and attachment &	ER &	Dissociative neutral recombination  (AB$^+$ + e$^-$ $\rightarrow$ A + B) or the emission of a photon\\
& &  (AB$^+$ + e$^-$ $\rightarrow$ AB + photon) or the attachment of  an electron (A + e$^-$ $\rightarrow$ A$^-$ + photon). \\
Third-body assisted association &	3-body &	Association reactions between two species, stabilized by collision with a third body (A + B + C $\rightarrow$ AB + C).\\
\enddata
\tablenotetext{a}{Charge exchange reactions have been defined as a separate type due to the particular microphysics behind this process.}
\end{deluxetable}

\clearpage

\begin{table*}
\begin{center}
\caption{The types of distributions and uncertainty factors used in KIDA.\label{unc}}
\begin{tabular}{llll}
\tableline\tableline
Type &	Distribution 	& $F_0$ meaning 	& $F_0$ unit\\
\tableline
norm &	Normal &	stdev$^{1}$: Pr$^2$($k_0 - F_0 \leq k \leq k_0 + F_0$) $\sim$ 68 \% &	as rate coefficient\\
unif &	Uniform &	half range: Pr$^2$($k_0 - F_0 \leq k \leq k_0 + F_0$) = 100 \%& 	as rate coefficient\\
logn &	Lognormal &	geometric stdev$^1$: Pr$^2$($k_0 / F_0 \leq k \leq k_0*F_0$) $\sim$ 68 \% &	no units\\
logu &	Loguniform &	geometric half range: Pr$^2$($k_0 / F_0 \leq k \leq k_0*F_0$) = 100 \% &	no units\\
\tableline
\end{tabular}
%% Any table notes must follow the \end{tabular} command.
\end{center}
$^1$ stdev = standard deviation\\
$^2$ Pr = Probability of assertion\\
\end{table*}

\begin{table*}
\begin{center}
\caption{Reactions involving grains \label{grain_reac}}
\begin{tabular}{lll}
\tableline\tableline
Reaction & A (cm$^{3}$ s$^{-1}$) & B \\
\tableline
C$^+$ +     GRAIN$^-$  $\rightarrow$        C  +     GRAIN0 &  $2.3\times 10^{-5}$ & 0.5 \\
Fe$^+$ +    GRAIN$^-$   $\rightarrow$       Fe  +   GRAIN0     & $1.1\times 10^{-5}$ & 0.5 \\
H$^+$    +  GRAIN$^-$    $\rightarrow$       H     +  GRAIN0      &         $7.9\times 10^{-5}$ & 0.5 \\          
He$^+$   +  GRAIN$^-$  $\rightarrow$         He    +  GRAIN0     &                  $3.9\times 10^{-5}$ & 0.5 \\
Mg$^+$  +   GRAIN$^-$  $\rightarrow$        Mg    +  GRAIN0      &                 $1.6\times 10^{-5}$ & 0.5 \\
N$^+$   +   GRAIN$^-$     $\rightarrow$      N   +    GRAIN0         &        $2.1\times 10^{-5}$ & 0.5 \\
Na$^+$  +   GRAIN$^-$    $\rightarrow$       Na  +    GRAIN0        &            $1.6\times 10^{-5}$ & 0.5 \\
O$^+$   +   GRAIN$^-$   $\rightarrow$        O   +    GRAIN0           &           $2.0\times 10^{-5}$ & 0.5 \\
S$^+$   +   GRAIN$^-$     $\rightarrow$     S    +   GRAIN0              &         $1.4\times 10^{-5}$ & 0.5 \\
Si$^+$   +  GRAIN$^-$     $\rightarrow$     Si   +   GRAIN0                 &       $1.5\times 10^{-5}$ & 0.5 \\
H$_3^+$ +    GRAIN$^-$    $\rightarrow$       H$_2$ +      H      + GRAIN0 &      $4.6\times 10^{-5}$ & 0.5 \\  
HCO$^+$  +  GRAIN$^-$    $\rightarrow$     H  +     CO  +    GRAIN0  &         $1.5\times 10^{-5}$ & 0.5 \\     
e$^-$ + GRAIN0 $\rightarrow$ GRAIN$^-$ & $3.4\times 10^{-3}$ & 0.5 \\
\tableline
\end{tabular}
\end{center}
\end{table*}%

\begin{table}
\begin{centering}
\begin{tabular}{ll}
\hline 
Distribution & Formula\tabularnewline
\hline 
Lognormal & $k(T)=\exp\left(\ln k_{0}(T)+\ln F(T)*N(0,1)\right)$\tabularnewline
Normal & $k(T)=k_{0}(T)+F(T)*N(0,1)$\tabularnewline
Uniform & $k(T)=k_{0}(T)+F(T)*2*(U(0,1)-0.5)$\tabularnewline
Loguniform & $k(T)=\exp\left(\ln k_{0}(T)+\ln F(T)*2*(U(0,1)-0.5)\right)$\tabularnewline
\hline 
\end{tabular}
\par\end{centering}

\caption{Generating random samples for MCUP. $U(0,1)$ is a standard
uniform random number generator (between 0 and 1); $N(0,1)$ is a
standard normal/gaussian random number generator (centered at 0;
variance 1). In systems with a temperature gradient, and where multiple
uncertain rate coefficients are managed simultaneously, care has to
be taken that the same random number is used for the whole temperature
range of a single reaction. \label{dist}}
\end{table}

%\begin{table}
%\begin{center}
%\caption{Model parameters for dense cloud chemistry. \label{cloudparam}}
%\begin{tabular}{ll}
%\tableline\tableline
%Parameter & Value \\
%\tableline
%Temperature & 10~K\\
%H density & $2\times 10^4$~cm$^{-3}$ \\
%Cosmic-ray ionization rate & $1.3\times 10^{-17}$~s$^{-1}$\\
%A$_{\rm v}$ & 30\\
%Initial abundances & from Wakelam et al. (2010)\\
%\tableline
%\end{tabular}
%\end{center}
%\end{table}%

\end{document}